\documentclass[aps,floats]{revtex4}
\usepackage{amsmath,amssymb}
\usepackage{graphicx,epsfig}

\begin{document}
\bibliographystyle {plain}

\def\oppropto{\mathop{\propto}} 
\def\opsimeq{\mathop{\simeq}}
\def\opoverderline{\mathop{\overline}}
\def\operarrow{\mathop{\longrightarrow}}
\def\opsim{\mathop{\sim}}

\def\fig#1#2{\includegraphics[height=#1]{#2}}
\def\figx#1#2{\includegraphics[width=#1]{#2}}


\title{ Directed polymer in a random medium of dimension $1+3$ : \\
multifractal properties at the localization/delocalization transition   } 


\author{ C\'ecile Monthus and Thomas Garel }
 \affiliation{Service de Physique Th\'{e}orique, CEA/DSM/SPhT\\
Unit\'e de recherche associ\'ee au CNRS\\
91191 Gif-sur-Yvette cedex, France}

\begin{abstract}

We consider the model of the directed polymer in a random medium of dimension $1+3$, and investigate its multifractal properties at the localization/delocalization transition. In close analogy with models of the quantum Anderson localization transition, where the multifractality of critical wavefunctions is well established, we analyse the statistics of the position weights $w_L(\vec r)$ of the end-point of the polymer of length $L$ via the moments $Y_q(L) = \sum_{\vec r} [w_L(\vec r)]^q$. We measure the generalized exponents $\tau(q)$ and $\tilde \tau(q)$ governing the decay of the typical values $Y^{typ}_q(L) = e^{\overline{\ln Y_q(L)}} \sim L^{-  \tau(q)} $ and disorder-averaged values $\overline{Y_q(L)} \sim L^{- \tilde \tau(q)} $ respectively. To understand the difference between these exponents, $ \tau(q) \neq  \tilde \tau(q)$ above some threshold $q>q_c \sim 2$, we compute the probability distributions of $y=Y_q(L)/Y^{typ}_q(L) $ over the samples : we find that these distributions becomes scale invariant with a power-law tail $1/y^{1+x_q}$. These results thus correspond to the Ever-Mirlin scenario [Phys. Rev. Lett. 84 , 3690 (2000)] for the statistics of Inverse Participation Ratios at the Anderson localization transitions. Finally, the finite-size scaling analysis in the critical region yields the correlation length exponent $\nu \sim 2$.

\end{abstract}

\maketitle

\section{Introduction}

The notion of multifractals is now important in various areas
of physics (see for instance \cite{halsey,Pal_Vul,Stan_Mea,Aha,Meakin,
harte,duplantier_houches} and references therein).
For classical systems with frozen disorder of interest here,
the idea that multifractality is present at
criticality has been mostly investigated for
correlation functions in two-dimensional diluted ferromagnets
\cite{Ludwig,Jac_Car,Ols_You,PCBI,Cha_Ber}, spin-glasses and random
field spin systems
\cite{Sourlas,Thi_Hil,Par_Sou}. For disordered quantum spin-chains,
the statistics of critical correlation functions is described by
``multiscaling'',
which is even stronger than multifractality \cite{multiscaling}.
For quantum localization models, the multifractality of the critical wavefunction
shows up through the statistics
of inverse participation ratios (I.P.R.s) \cite{Weg,Cas_Pel}.
Many results are now available for
the Anderson localization transition in $d=3$
\cite{Schreiber,Terao,Mirlin2002,Mirlin2006},  
the integer quantum Hall transition \cite{Jan,Huck}
 Dirac fermions in a random magnetic field \cite{Cha_etcie},
and power-law random banded matrices \cite{Mirlin_Evers}.
Connections have been also established
 with the scaling properties of the correlation
functions \cite{Pook} and with the time evolution of
wave packets \cite{timeexponents}.
More recently, it was realized that typical and disorder-averaged 
I.P.R.s can actually lead to two different
multifractal spectra as a consequence of the broadness of their probability distributions
 \cite{Mirlin_Evers,Mirlin2002,averaged-typical}.
  
In this paper, we consider the localization/delocalization transition
of the directed polymer in a random medium of dimension $1+3$
(see the review \cite{Hal_Zha} and references therein),
to investigate whether some multifractality in present at criticality,
in analogy with the quantum localization models quoted above.
 
The paper is organized as follows. In Section \ref{mod}, we introduce the directed polymer model 
and the observables displaying multifractal behavior at criticality.
We then describe our numerical results concerning  the generalized dimensions $D(q)$ and $\tilde D(q)$ 
that govern typical and averaged values (Section \ref{dq}),
 the singularity spectrum (Section  \ref{falpha}), and the probability distributions
 over the samples ( Section \ref{histoyq}). Finally, we present in
Section \ref{fss}  the finite-size scaling analysis
in the critical region. Section \ref{conclusion} contains our conclusions.

\section{ Models and observables }

\label{mod}

\subsection{ Model definition}

The random bond directed polymer
 model \cite{Hal_Zha} is defined by the 
following recursion relation for the partition function
on the cubic lattice in $d=3$
\begin{eqnarray}
\label{DP1}
Z_{t+1} (\vec r) =  \sum_{j=1}^{2d}
 e^{-\beta \epsilon_t(\vec r+\vec e_j,\vec r)} Z_{t} (\vec r+\vec e_j)
\label{transfer}
\end{eqnarray}
The bond energies $\epsilon_t(\vec r+\vec e_j,\vec r) $
are random independent variables drawn from the Gaussian
distribution 
\begin{eqnarray}
\rho (\epsilon) = \frac{1}{\sqrt{2\pi} } e^{- \frac{\epsilon^2}{2} }
\end{eqnarray}
In this paper, we consider the following boundary conditions.
 The first monomer is fixed at $\vec r = \vec 0$, 
i.e. the initial condition of the recurrence of Eq. (\ref{DP1}) reads
\begin{eqnarray}
Z_{t=0} (\vec r) = \delta_{\vec r, \vec 0}
\end{eqnarray}
The last monomer is free, i.e. the full partition function of the polymer of length $L$  
is then obtained by summing over all possible positions $\vec r$
at $t=L$
\begin{eqnarray}
Z_L^{tot} = \sum_{\vec r} Z_L(\vec r)  
\label{ztot}
\end{eqnarray}

The phase diagram of this directed polymer model as a function of space dimension $d$ is the
following \cite{Hal_Zha}. In dimension $d \leq 2$, there is no free phase,
i.e. any initial disorder drives the polymer into a strong disorder phase,
where the 
order parameter is an `overlap' \cite{Der_Spo,Mez,Car_Hu,Com,Var}.
In dimension, $d>2$, 
there exists a phase transition between
the low temperature disorder dominated phase
and a free phase at high temperature  \cite{Imb_Spe,Coo_Der,Ki_Br_Mo}.
 This phase transition 
has been studied exactly on a Cayley tree \cite{Der_Spo}.
  In finite dimensions, bounds on the critical temperature
$T_c$ have been derived \cite{Coo_Der,Der_Gol,Der_Eva} :
$T_0(d) \le T_c \le T_2(d)$.
The upper bound $T_2(d)$ 
corresponds to the temperature above which the ratio
$\overline{Z_L^2}/(\overline{Z_L})^2$ remains finite as $L \to
\infty$. The lower bound $T_0$ corresponds to the temperature below which
the annealed entropy becomes negative.
On the Cayley tree, the critical temperature $T_c$ coincides
with $T_0$ \cite{Der_Spo}.
In finite dimensions however, we have argued in \cite{DPcritidroplet}
that $T_c$ coincides with $T_2$, and our recent numerical simulations \cite{DPtransi3d}
are in agreement with the numerical value given in \cite{Der_Gol} for
$T_2(d=3)$ : 
\begin{eqnarray}
T_c= T_2(d=3) =0.790..
\label{tc3d}
\end{eqnarray}

The numerical results given below have been obtained using polymers of various lengths $L$,
with corresponding numbers $n_s(L)$ of disordered samples with
\begin{eqnarray}
L && =6,12,18,24,36,48,60,72,84,96 \\
n_s(L) && = 10^8, 10^7, 2.10^6, 8.10^5, 2.10^5, 5. 10^4, 3. 10^4, 2.10^4
4.10^4, 2.10^4
\label{nume}
\end{eqnarray}

In the following, $\overline{A}$ denotes the average of $A$ over the
disorder samples. We also define $(\Delta A)^2= \overline{A^2}-{\overline{A}}^2$.

\subsection{ Notion of multifractal statistics at criticality }

In this paper, we  will focus on the statistical properties 
of the weights   
\begin{eqnarray}
w_L (\vec r) = \frac{ Z_L(\vec r) }{ Z_L^{tot}} 
\label{defw}
\end{eqnarray}
normalized to (Eq. \ref{ztot})
\begin{eqnarray}
\sum_{\vec r} w_L (\vec r) = 1
\label{norma}
\end{eqnarray}
and study whether they present some multifractal properties at criticality.
We thus consider the following moments of arbitrary order $q$
\begin{equation}
Y_q(L)  = \sum_{ \vec r} w_L^q(\vec r)
\label{ykdef}
\end{equation}
that are the dynamical analogs of the Inverse Participation Ratios (I.P.R.s)
in quantum localization models \cite{Mirlin_Evers}
\begin{equation}
P_q(L)  = \int_{L^d} d^d { \vec r}  \vert \psi (\vec r) \vert^{2q}
\label{ipr}
\end{equation}

As a consequence of the normalization of Eq. \ref{norma},
one has the identity $Y_{q=1}(L)=1$.
The localization/delocalization transition can 
be characterized by the asymptotic behavior 
in the limit $L \to \infty$ of the $Y_q(L)$ for $q>1$.
In the localized phase $T<T_c$ 
these moments $Y_q(L)$ converge to finite values
\begin{equation}
Y_q(L=\infty) >0 \ \ {\rm for } \ \ T<T_c
\label{loc}
\end{equation}
In the delocalized phase, the spreading of the polymer 
involves the Brownian exponent $\zeta=1/2$,
and, space being of dimension $d=3$, the decay
of the moments follows the scaling
\begin{equation}
Y_q(L) \simeq L^{ -(q-1) d \zeta} =
 L^{ -(q-1) \frac{3}{2} } \ \ {\rm for } \ \ T>T_c
\label{deloc}
\end{equation}
Exactly at criticality, the typical decay of the $Y_q(L)$
defines a series of generalized exponents $\tau(q)=(q-1) D(q) $
\begin{equation}
Y_q^{typ}(L) \equiv e^{ \overline{ \ln Y_q(L)} }  \vert_{T=T_c} \simeq 
L^{ - \tau(q) } = L^{ -(q-1) D (q) } 
\label{tctyp}
\end{equation}
The notion of multifractality corresponds to the case where $D(q)$
depends on $q$, whereas monofractality corresponds to $D(q)=cst$
as in Eq. (\ref{deloc}).
The exponents $D(q)$ represent generalized dimensions \cite{halsey} :
$D(0)$ represent the dimension of the support of the measure,
here it is simply given by the space dimension $D(0)=d=3$;
 $D(1)$ is usually called the information dimension \cite{halsey} ,
since it describes the behavior of 
the 'information' entropy
\begin{equation}
s_L \equiv - \sum_{ \vec r } w_L(\vec r) \ln  w_L(\vec r)
= - \partial_q Y_q(L) \vert_{q=1} \simeq  D(1) \ln L
\label{entropy}
\end{equation}
Finally $D(2)$ is called the correlation dimension \cite{halsey}
and describes the decay of
\begin{equation}
Y_2^{typ}(L) \equiv e^{ \overline{ \ln Y_2(L)} }  \vert_{T=T_c} \simeq 
 L^{ - D (2) } 
\label{y2d2}
\end{equation}

In the multifractal formalism, the singularity spectrum $f(\alpha)$
is given by the Legendre transform of $\tau(q)$ \cite{halsey}
via the standard formulas
\begin{eqnarray}
 q && =f'(\alpha) \\
 \tau(q) && = \alpha q - f(\alpha)
\label{legendre}
\end{eqnarray}
The physical meaning of $f(\alpha)$ is that the number ${\cal N}_L(\alpha)$
of points $\vec r$ where the weight $w_L(\vec r)$
scales as $L^{-\alpha}$ typically behaves as 
\begin{eqnarray}
{\cal N}_L(\alpha) \propto L^{f(\alpha)}
\label{nlalpha}
\end{eqnarray}
So the Legendre transform of Eq. (\ref{legendre}) corresponds to
the saddle-point calculus in $\alpha$ of the following expression
\begin{equation}
Y_q^{typ}(L) \sim \int d\alpha \ L^{f(\alpha)} \ L^{- q \alpha} 
\label{saddle}
\end{equation}
The general properties of 
the singularity spectrum $f(\alpha)$ are as follows \cite{halsey} :
it is positive $f(\alpha) \geq 0$
 on an interval $[\alpha_{min},\alpha_{max}]$
where $\alpha_{min}=D(q=+\infty)$ is the minimal singularity exponent
and $\alpha_{max}=D(q=-\infty)$ is the maximal singularity exponent.
It is concave $f''(\alpha)<0$. 
It has a single maximum at some value $\alpha_0$ where
$f(\alpha_0)=D(q=0)$ (so here $f(\alpha_0)= 3$), 
and 
contains the point $\alpha_1=f(\alpha_1)=D(1)$.

Following \cite{halsey}, many authors consider that
the singularity spectrum has a meaning only for $f(\alpha) \geq 0$
\cite{Schreiber,Terao,Jan,Huck,Cha_etcie}. However, when
multifractality arises in random systems, disorder-averaged values
may involve other generalized exponents \cite{mandelbrot,Chh_neg,Jen_neg,has_dup}
than the typical values (see Eq. \ref{tctyp}).
In quantum localization transitions, these exponents were 
denoted by  $\tilde \tau(q)=
(q-1) \tilde D(q) $ in \cite{Mirlin_Evers,averaged-typical}
\begin{equation}
\overline{ Y_q(L)} \vert_{T=T_c} \simeq
 L^{ -\tilde \tau(q) } = L^{ -(q-1) \tilde D (q) } 
\label{tcav}
\end{equation}
For these disorder averaged values, the corresponding singularity
spectrum
$\tilde f(\alpha)$ may become negative $\tilde f(\alpha)<0$ 
\cite{mandelbrot,Chh_neg,Jen_neg,has_dup,Mirlin_Evers,averaged-typical}
to describe rare events (cf Eq. \ref{nlalpha}).
The difference between the two generalized exponents sets $D(q)$ and
$\tilde D(q)$ associated to typical and averaged values 
has for origin the broad
distributions at criticality \cite{Mirlin_Evers,Mirlin2002}
as we now describe.

\subsection{ Probability distributions of the $ Y_q(L)$ }

The scenario proposed in \cite{Mirlin_Evers,Mirlin2002} in the context of
quantum localization models 
is as follows : the probability distribution of the logarithm
of the Inverse Participation Ratios of Eq. \ref{ipr} becomes scale invariant
around its typical value
\cite{Mirlin_Evers,Mirlin2002}, i.e.
\begin{equation}
\ln P_q(L) = \overline{\ln P_q(L) } + u
\label{lnpq}
\end{equation}
where $u$ remains a random variable of order $O(1)$ 
in the limit $L \to \infty$. According to \cite{Mirlin_Evers}
the probability distribution
$G_L(u)$ generically develops an exponential tail
\begin{equation}
G_{\infty}(u) \opsimeq_{u \to \infty} e^{ - x_q u}
\label{utail}
\end{equation}
As a consequence, the ratio $y =  P_q(L)/P_q^{typ}(L)=e^{u}$
with respect to the typical value $P_q^{typ}(L)=e^{\overline{\ln P_q(L)}}$
presents the power-law decay
\begin{equation}
\Pi \left( y \equiv \frac{P_q(L)}{P_q^{typ}(L)} \right) 
\oppropto_{ y \to \infty } \frac{1}{y^{1+x_q}}
\label{tailp}
\end{equation}
The conclusions of \cite{Mirlin_Evers,Mirlin2002} are then as follows :
for $q$ small enough $q<q_c$, the exponent satisfies $x_q>1$,
and the corresponding generalized dimensions coincide $ \tilde D (q)= D(q)$.
However, for larger values $q>q_c$, the exponent $x(q)$ may become smaller
$x(q)<1$, and then the corresponding generalized dimensions differ
 $ \tilde D (q) \neq D(q)$ : the decay of the averaged value 
$ \overline{P_q(L)}$ is then determined by the finite-size cut-off
of the power-law tail. In this case, the averaged values are not
representative but are governed by rare events.

In this paper, we show that this scenario for the I.P.R.'s statistics at Anderson transitions
describes well our data for the directed polymer at criticality.

\section{ Results for the generalized exponents $D(q)$ and $\tilde D(q)$}

\label{dq}

\begin{figure}[htbp]
\includegraphics[height=6cm]{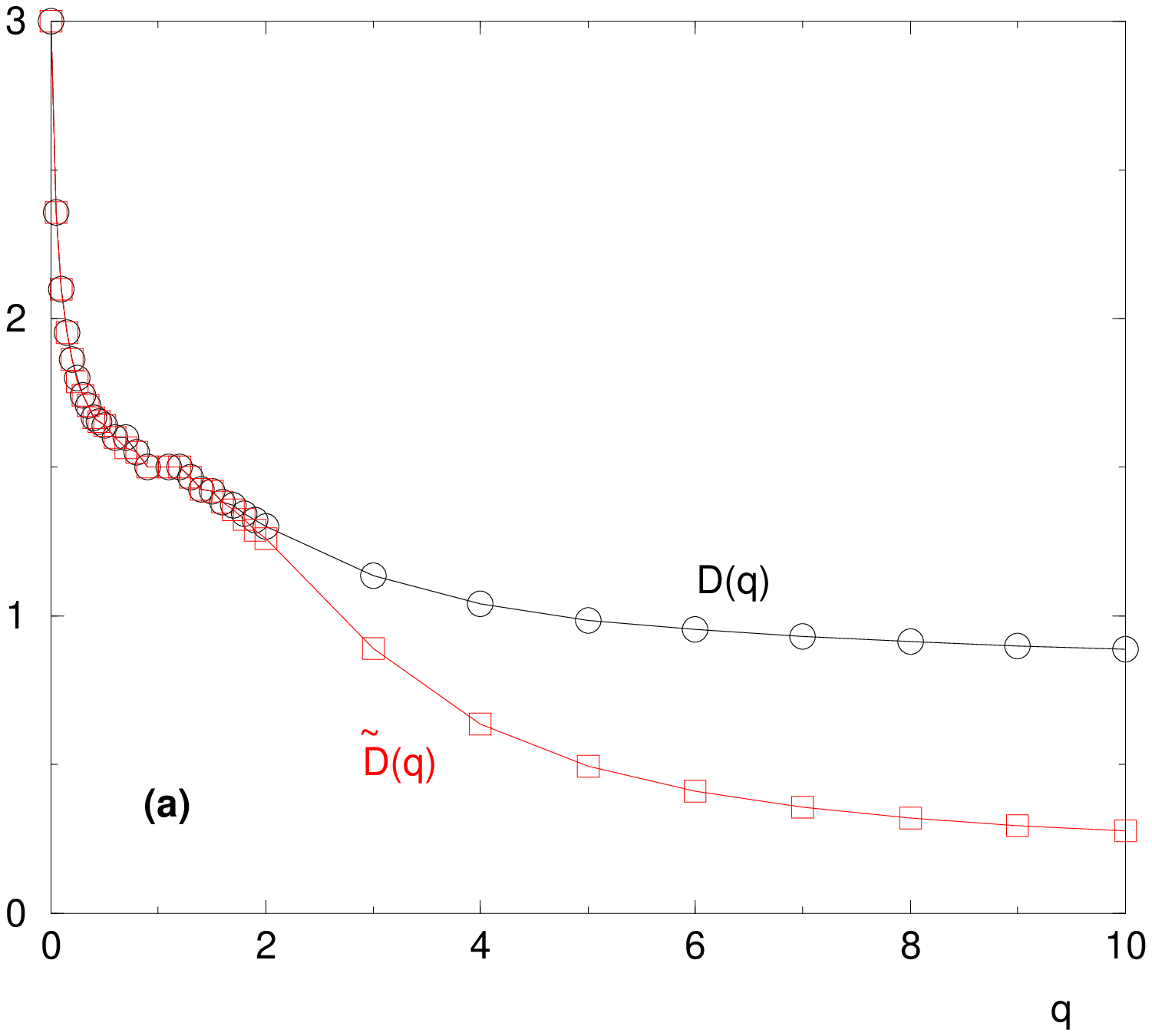}
\hspace{1cm}
 \includegraphics[height=6cm]{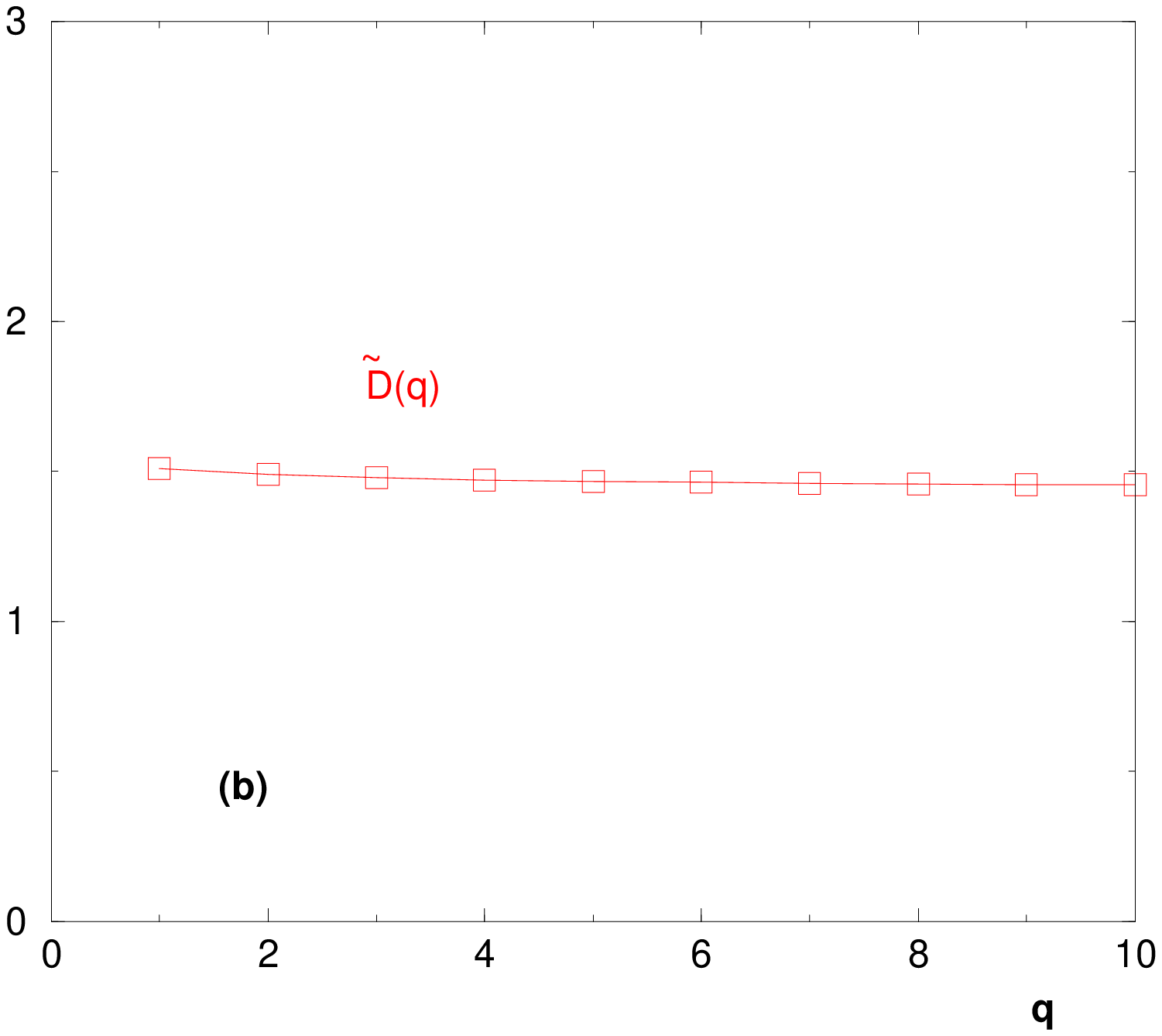}
\caption{(Color online)  
(a) Multifractality at criticality ($T=0.79$) : generalized dimensions $D(q)$ ($\bigcirc$)
and  $\tilde D(q)$ ($\square$) 
associated to typical and disorder averaged values
(Eq. \ref{tctyp} and \ref{tcav}) 
(b) Monofractality in the high-temperature phase ($T=2.$): $D(q)=\tilde D(q)=\frac{3}{2}$ (see Eq. \ref{deloc})}
\label{figexposantdk}
\end{figure}

We show on Fig. \ref{figexposantdk} (a) our results for the generalized exponents $D(q)$ and  $\tilde D(q)$ 
governing the decay of typical and disorder averaged values
(Eqs. \ref{tctyp} and \ref{tcav}).
In agreement with the scenario proposed in
\cite{Mirlin_Evers,Mirlin2002} for Anderson transitions, we find that 
there exists a threshold $q_c$, of order $q_c \sim 2$ here, such that
\begin{eqnarray}
D(q) && =\tilde D(q) \ \ {\rm for } \ \ q<q_c \\ 
D(q) && > \tilde D(q) \ \ {\rm for } \ \ q>q_c 
\label{qc}
\end{eqnarray}
In particular, for $q=1$, the information dimension of Eq.(\ref{entropy}) is 
\begin{eqnarray}
D(1)  =\tilde D(1) \sim 1.5
\label{d1}
\end{eqnarray}
and corresponds to the monofractal dimension $D_{T>T_c}(q) = 3/2$
of the delocalized phase (see Eq. \ref{deloc}), as numerically checked 
on Fig. \ref{figexposantdk} (b).

For $q=2$, the correlation dimension $D(2)$ defined in Eq. (\ref{y2d2})
is found to be of order
\begin{eqnarray}
D(2)  \sim \tilde D(2) \sim 1.3
\label{d2}
\end{eqnarray}

For $q \geq 3$, the values for $D(q)$ and  $\tilde D(q)$ are clearly
different,
in particular for $q=3$
\begin{eqnarray}
D(3)  && \sim 1.1  \\
\tilde D(3) && \sim 0.9
\label{d3}
\end{eqnarray}
and for $q=10$
\begin{eqnarray}
D(10)  && \sim 0.9  \\
\tilde D(10) && \sim 0.3
\label{d10}
\end{eqnarray}
The limit $q \to \infty$ will be discussed more precisely below (see
Eq. \ref{alphamin})

Finally, for $q<0$, we find that $Y_q^{typ}(L)$ and
$\overline{Y_q(L)}$ diverge more rapidly than the power-laws of
Eqs. \ref{tctyp} and \ref{tcav}, i.e.
\begin{eqnarray}
D(q<0)  && = +\infty  
\label{dneg}
\end{eqnarray}
This can be understood in the delocalized phase $T>T_c$ where
it is also true, since for the free Gaussian probability
$w_{T=\infty}(\vec r) \sim e^{- (\vec r)^2/L}/L^{3/2}$ at $T=\infty$
leads to exponential divergence of $Y_q$ in the negative region $q<0$.
This finding for our 'dynamical' model is thus very natural,
but is a major difference with the Anderson localization cases
where the generalized exponents $D(q)$ are finite for $q<0$.

\section{ Results for the singularity spectrum $f(\alpha)$ }

\label{falpha}

\begin{figure}[htbp]
\includegraphics[height=6cm]{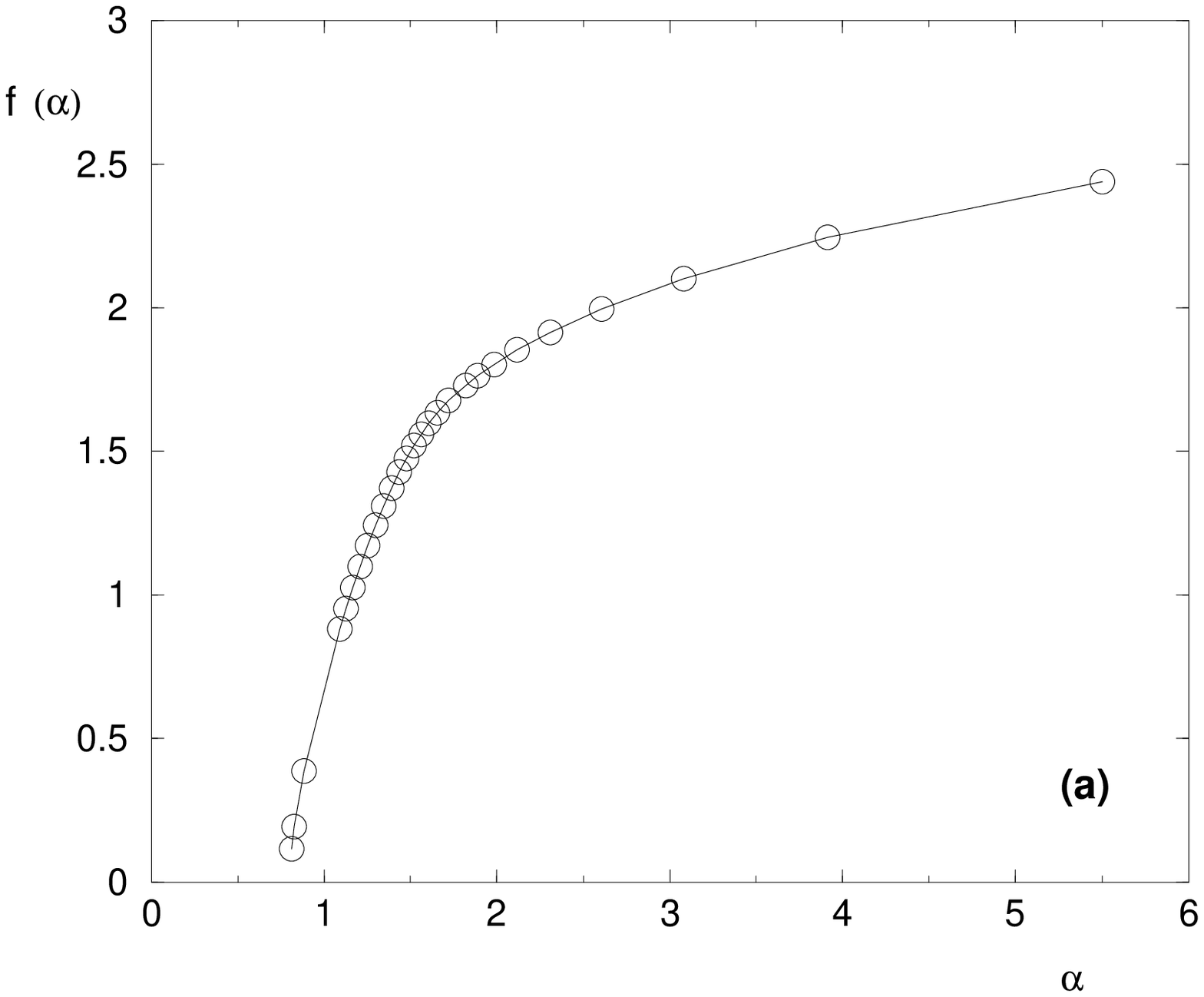}
\hspace{1cm}
 \includegraphics[height=6cm]{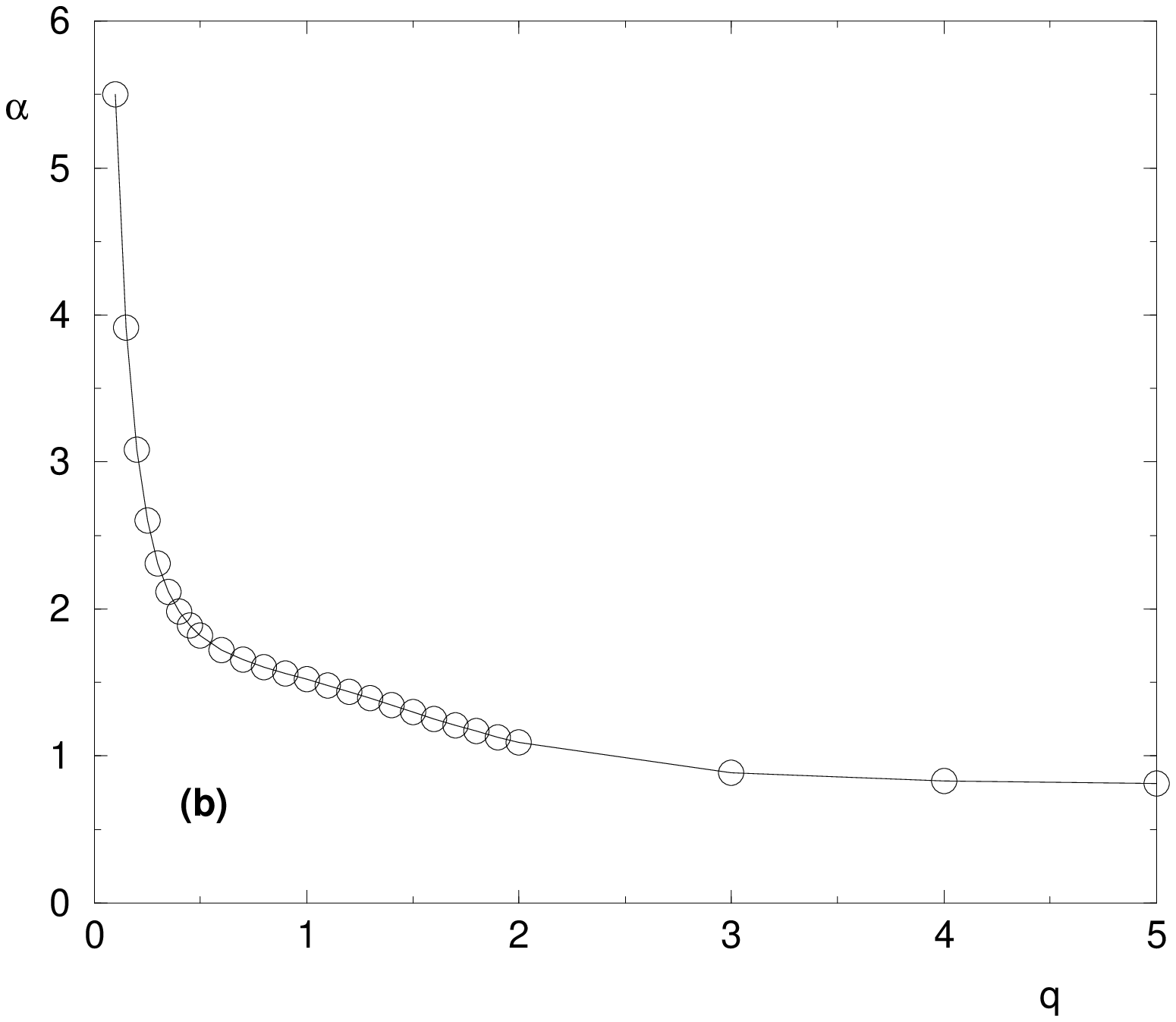}
\caption{ 
(a)   Singularity spectrum $f(\alpha)$ : starting at some
$\alpha_{min}=D(q=+\infty)\sim 0.77$ 
where $f(\alpha_{min})=0$, it is tangent to the diagonal 
$\alpha=f(\alpha)$ at
 $\alpha_1=D(1)\sim 1.5$ and asymptotically
goes to $f(+\infty)=D(0)=3$.
(b) Corresponding curve $\alpha(q)$ : diverging at $q \to 0$, it goes
through the point $\alpha(q=1)=D(1) \sim 1.5$ and  tends to $
\alpha_{min} =D(q=+\infty)$ as $ q \to \infty$.  }
\label{figspectre}
\end{figure}

To measure the singularity spectrum $f(\alpha)$, we have followed
 the method of the q-measures proposed in \cite{Chh}.
As explained above, in the negative region $q<0$, our model
does not lead to power-law (see Eq. \ref{dneg}).
 As a consequence, the maximum $(\alpha_0,f(\alpha_0))$ of the curve
$f(\alpha)$
which is at finite distance in Anderson localization models,
is rejected towards infinity in our case
\begin{eqnarray}
\alpha_0  && =+\infty   \\
f(\alpha_0)&& =D(q=0)=3 
\label{alpha0}
\end{eqnarray}

Our result for the curve $f(\alpha)$ are shown on
Fig. \ref{figspectre} (a) :
 it begins at some $\alpha_{min}=D(q=+\infty)$ 
where $f(\alpha_{min})=0$, it is tangent to the diagonal at
$\alpha_1=D(1) \sim 1.5 $ and asymptotically 
goes to $f(+\infty)=D(0)=3$.

The q-values used in our computations, and the corresponding
$\alpha(q)$ (see Eq. \ref{saddle}) are shown on Fig.  \ref{figspectre}
(b).

To measure better the minimal exponent $\alpha_{min}=D(q=+\infty)$ where
$f(\alpha)$ vanishes $f(\alpha_{min}=0)=0$, we have studied the
statistics of the maximal weight in each sample (Eq. \ref{defw})
\begin{eqnarray}
w_L^{max} = max_{\vec r} \left[ w_L(\vec r) \right]
\label{wmax}
\end{eqnarray}
The disorder-averaged value of its logarithm gives
\begin{eqnarray}
\overline{ \ln w_L^{max} } \sim - 0.77 \ln L
\label{lnwmax}
\end{eqnarray}
The first moment involves a similar value
\begin{eqnarray}
\overline{  w_L } \sim L^{-0.75}
\label{wmaxav}
\end{eqnarray}
so our conclusion is that the minimal exponent in a typical sample is around
\begin{eqnarray}
\alpha_{min}=D(q=+\infty) \sim 0.77
\label{alphamin}
\end{eqnarray}

\section{ Results for probability distributions }

\label{histoyq}

\subsection{ Probability distributions of the $ Y_q(L)$ }

\begin{figure}[htbp]
\includegraphics[height=6cm]{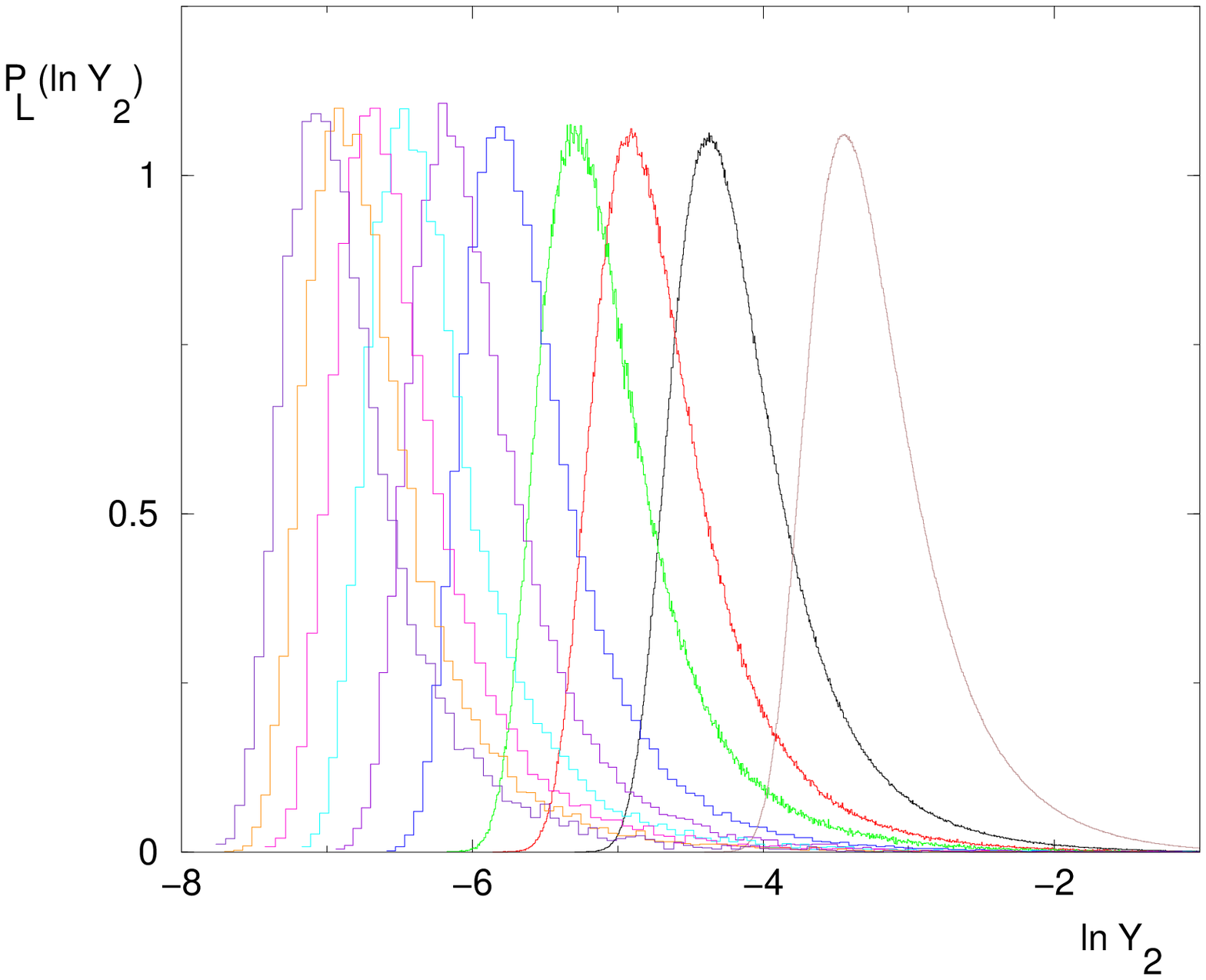}
\hspace{1cm}
\includegraphics[height=6cm]{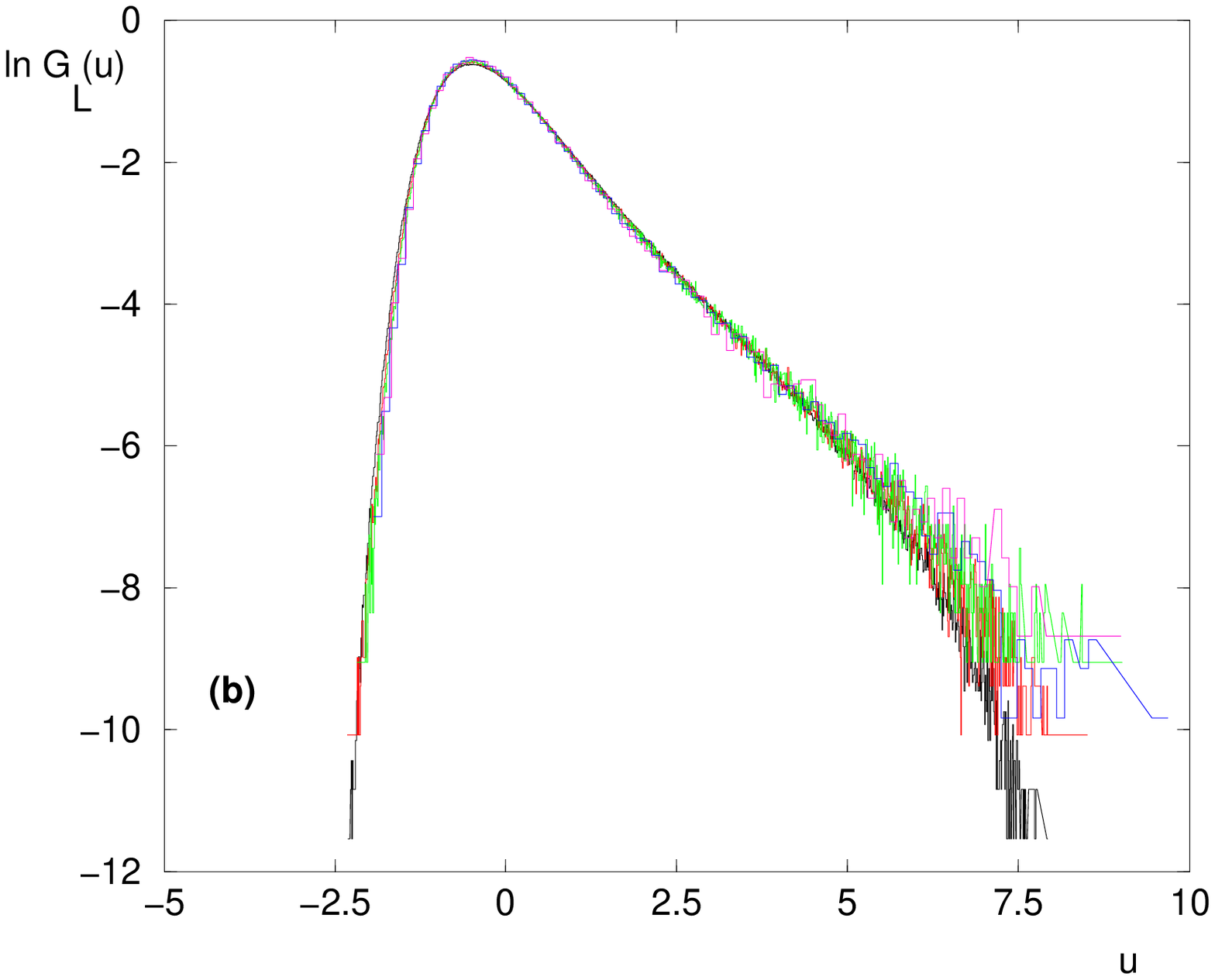}
\caption{(Color online) Histogram of $\ln Y_2(L)$ at criticality ($T_c=0.79$)
(a) Probability distribution $P_L(\ln Y_2)$ for
$L=6,12,18,24,36,48,60,72,84,96$ 
(b) Rescaled distributions $\ln G_L(u=\ln Y_2(L) - \overline{\ln Y_2(L)})$
 : the exponential tail of Eq.(\ref{utailbis})
is clearly visible, the corresponding slope being $x_2 \sim 1.$ }
\label{fighistolny2}
\end{figure}

\begin{figure}[htbp]
\includegraphics[height=6cm]{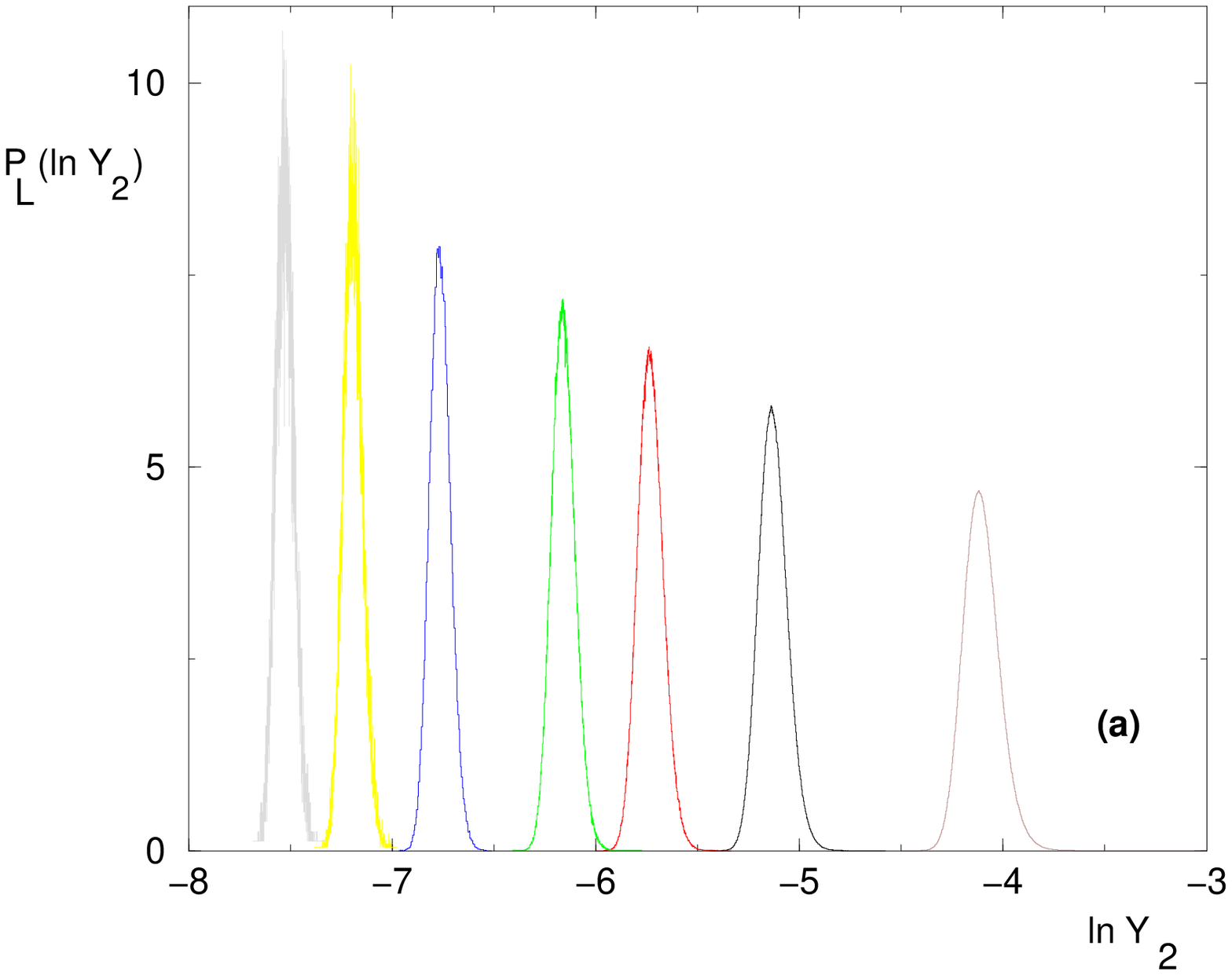}
\hspace{1cm}
\includegraphics[height=6cm]{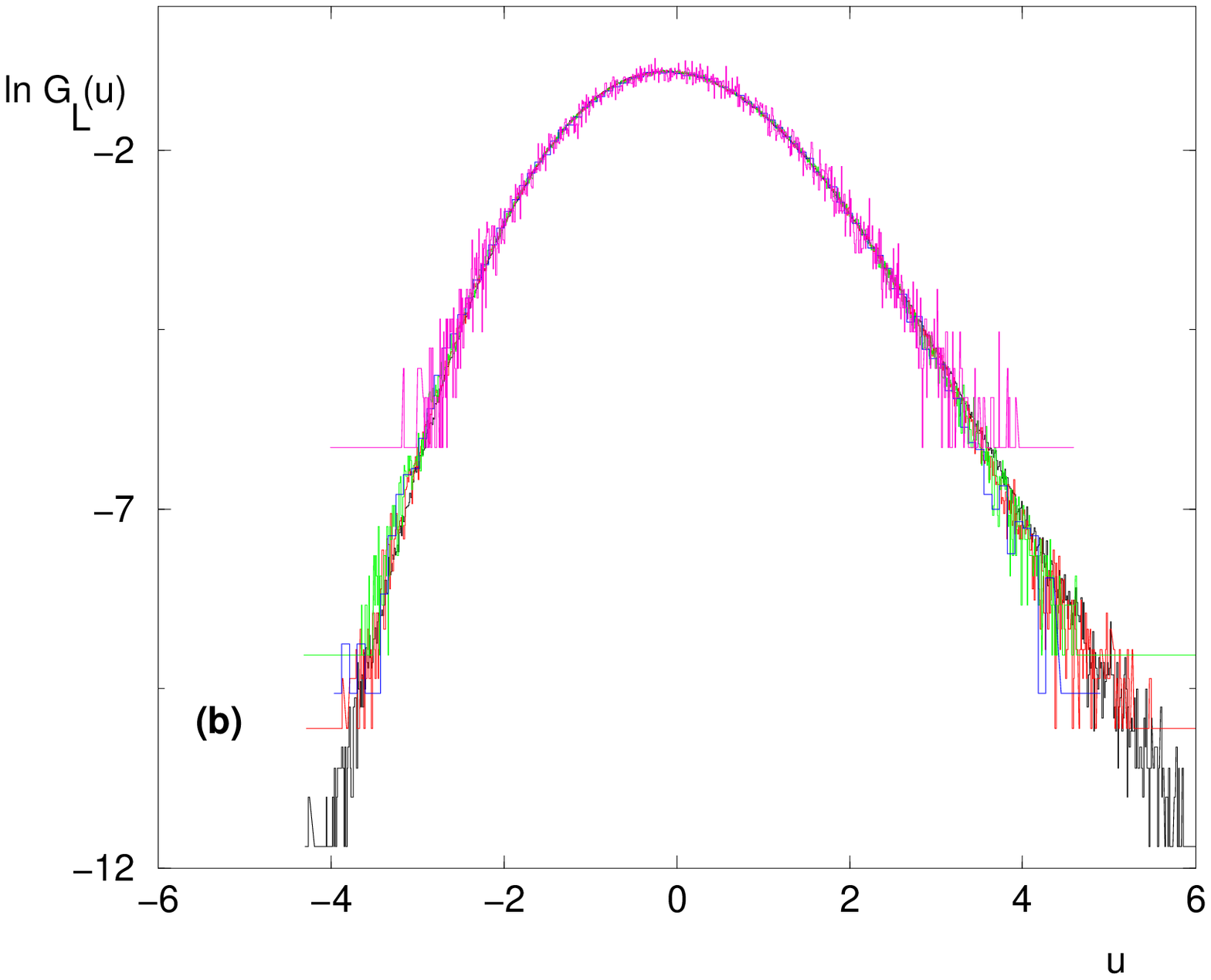}
\caption{(Color online) Histogram of $\ln Y_2(L)$ in the
high-temperature phase ( $T=2$ )
(a) Probability distribution of $P_L(\ln Y_2)$ pour $L=6,12,18,24,36,48,60$
(b) The distributions $G_L(u)$ of the rescaled variable 
$u=(\ln Y_2(L) -\overline{\ln Y_2(L)})/[\Delta \ln Y_2(L)]$ become Gaussian.
 }
\label{fighistolny2t2}
\end{figure}

To understand the difference between the generalized exponents
associated to typical and averaged values (Eq. \ref{qc}), we now consider
the probability distributions of $Y_q(L)$ over the samples.
Our results for the histograms of $\ln Y_2(L)$ for various $L$
at criticality are shown on Fig. \ref{fighistolny2} (a) .
Remarkably, as $L$ grows, this distribution simply shifts along the
$x$-axis with a fixed shape, as also found in
\cite{Mirlin_Evers,Mirlin2002}
for I.P.R.s at Anderson transitions.
As in Eq. \ref{lnpq},
we may therefore write
\begin{equation}
\ln Y_q(L) = \overline{\ln Y_q(L) } + u
\label{lnyq}
\end{equation}
where $u$ is a random variable of order $O(1)$ 
in the limit $L \to \infty$. 
The probability distribution $G_2(u)$ of $u=\ln Y_2(L) - \overline{\ln Y_2(L) }
$ is shown on Fig. \ref{fighistolny2} (b) for various $L$.
It clearly develops an exponential tail as $u \to \infty$
\begin{equation}
G_{L \to \infty}(u) \opsimeq_{u \to \infty} e^{ - x_q u}
\label{utailbis}
\end{equation}
As stressed in \cite{Mirlin_Evers,Mirlin2002},
the ratio $y =  Y_q(L)/Y_q^{typ}(L)=e^{u}$
then
presents the power-law decay
\begin{equation}
\Pi \left( y \equiv \frac{Y_q(L)}{Y_q^{typ}(L)} \right) 
\oppropto_{ y \to \infty } \frac{1}{y^{1+x_q}}
\label{tail}
\end{equation}
Whenever $x_q<1$, 
 the corresponding generalized dimensions differ
 $ \tilde D (q) \neq D(q)$ (Eq. \ref{qc}) : the decay of the averaged value 
$ \overline{Y_q(L)}$ is then determined by the finite-size cut-off
of the power-law tail.
Our results for the histograms of $Y_q(L)$ for $q=3,4,..$
are similar to the results shown for $q=2$ on Fig.  \ref{fighistolny2}.

For comparison, we show on Fig. \ref{fighistolny2t2} (a)
the histogram of $\ln Y_2(L)$ in the delocalized phase at $T=2$ :
as $L$ grows, the width shrinks around the averaged value
$\overline{\ln Y_2(L)} \sim - (3/2) \ln L$.
The corresponding rescaled distribution shown on Fig.
\ref{fighistolny2t2} (b) tends towards the Gaussian distribution.

\subsection{ Probability distributions of the last-monomer entropy
$s_L= - \sum_{ \vec r } w_L(\vec r) \ln  w_L(\vec r) $  }

\begin{figure}[htbp]
\includegraphics[height=6cm]{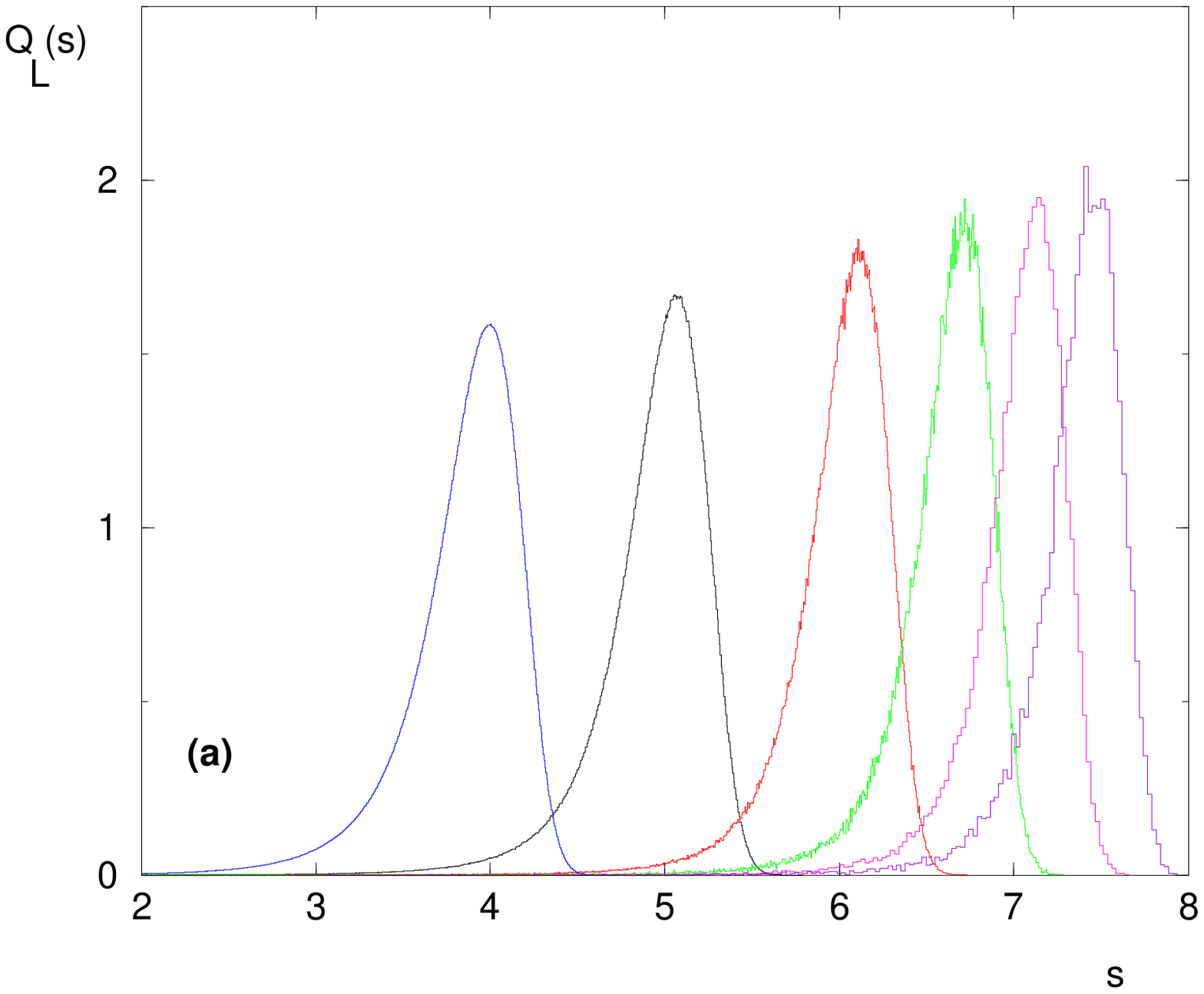}
\hspace{1cm}
\includegraphics[height=6cm]{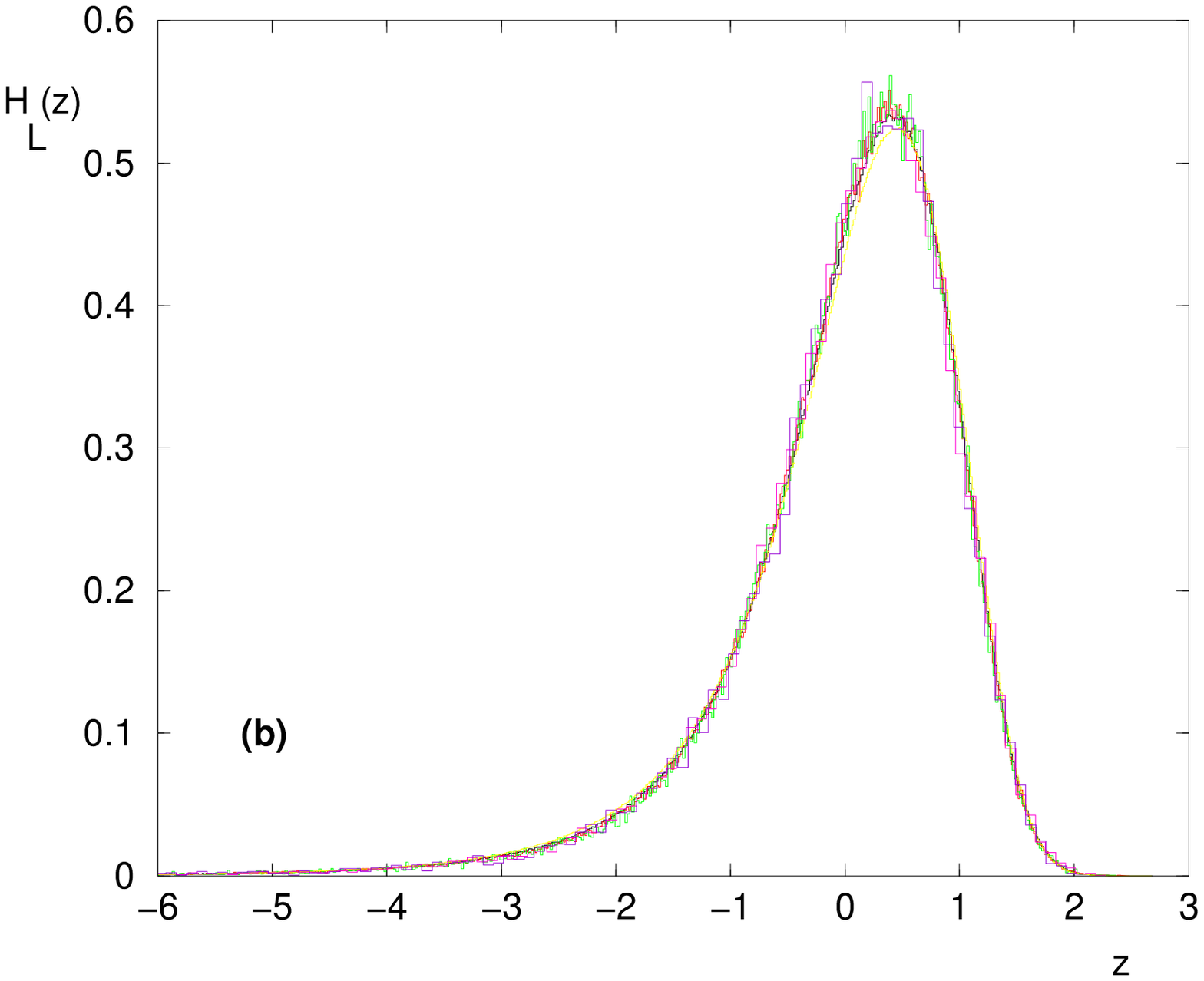}
\caption{(Color online) Histogram of the entropy $s_L= - \sum_{
\vec r } w_L(\vec r) \ln  w_L(\vec r)$ at criticality ($T=0.79$)
(a) Probability distribution $Q_L(s)$  for $L=6,12,24,36,48,60$ 
(b)The distributions $H_L(z)$ of the rescaled variable
$z=\frac{s_l-\overline{s_L}}{\Delta s_L}$ are clearly asymmetric.  }
\label{fig3dstc}
\end{figure}

\begin{figure}[htbp]
\includegraphics[height=6cm]{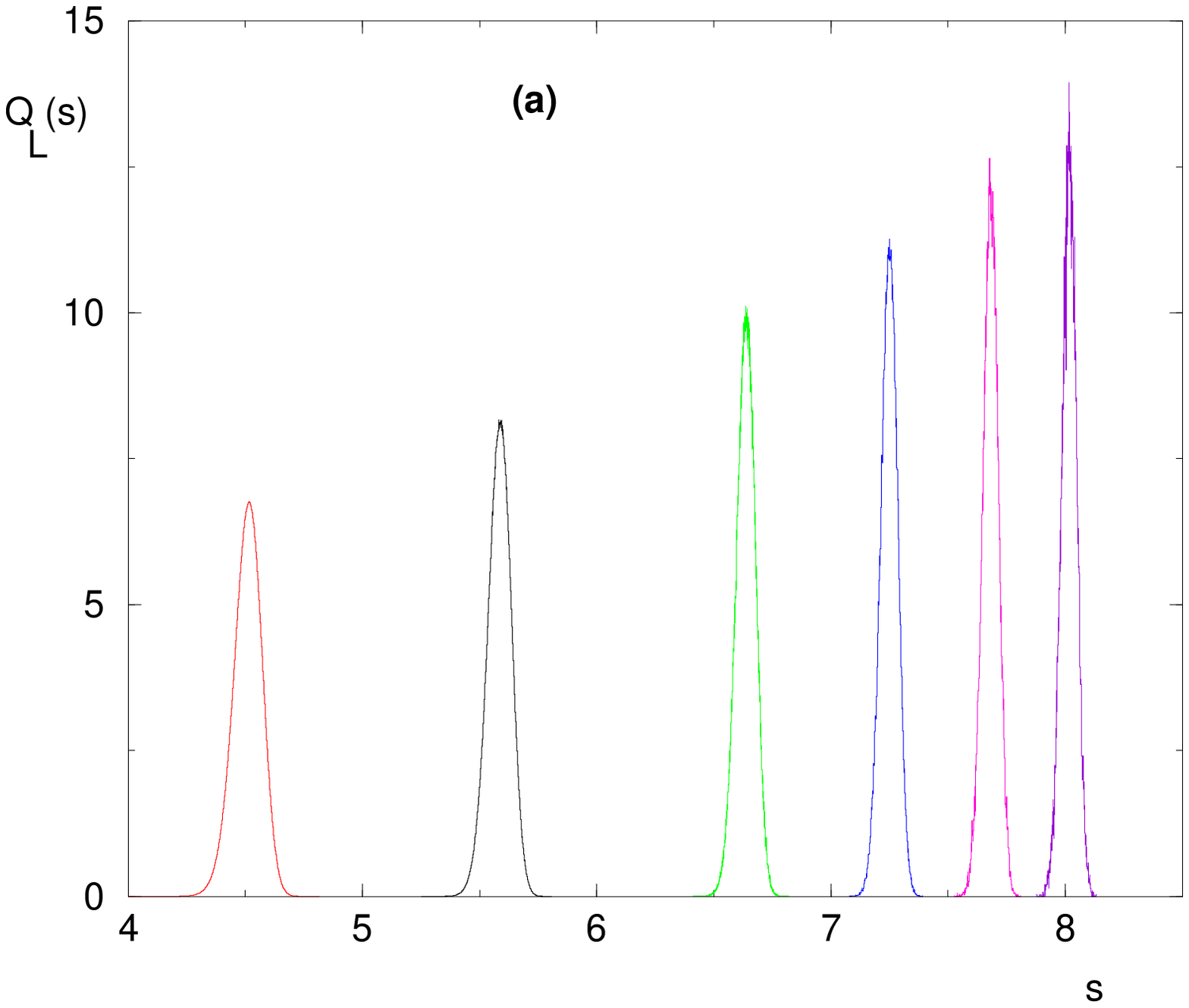}
\hspace{1cm}
\includegraphics[height=6cm]{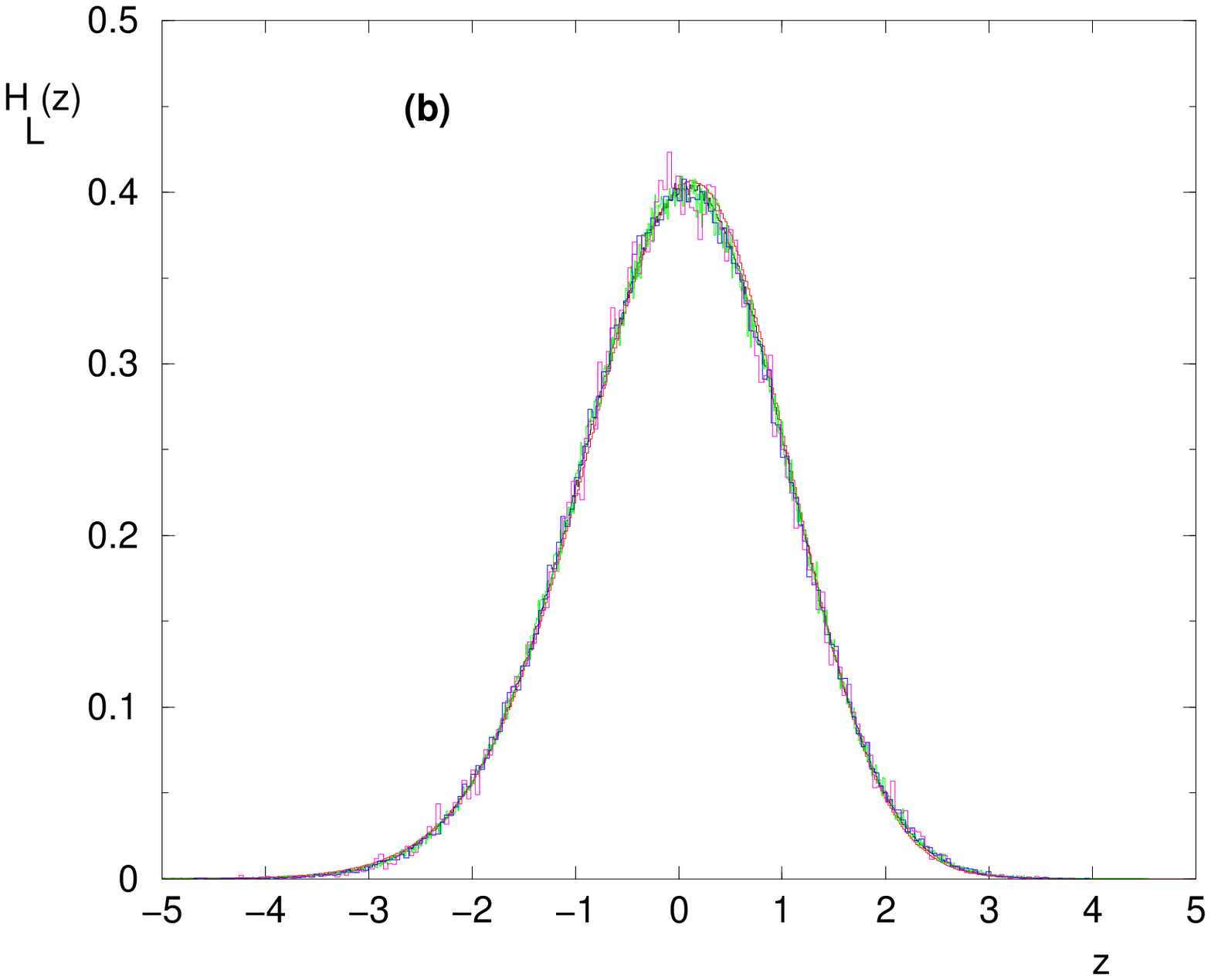}
\caption{(Color online) Histogram of the entropy $s_L= 
- \sum_{ \vec r } w_L(\vec r) \ln  w_L(\vec r)$ in the high
temperature phase ($T=2$)
(a)Probability distribution $Q_L(s)$ for $L=6,12,24,36,48,60$ 
(b) The distributions $H_L(z)$ of the rescaled variable
$z=\frac{s_l-\overline{s_L}}{\Delta s_L}$ become Gaussian. }
\label{fig3dst2}
\end{figure}

Since the last-monomer entropy  $s_L$ is closely related to the
$Y_q(L)$  (Eq. \ref{entropy}), we have also computed its histogram over the samples
both at criticality
(Fig \ref{fig3dstc} ) and in the delocalized phase at $T=2>T_c$ (Fig. \ref{fig3dst2}).
Again, the rescaled distribution is Gaussian for $T>T_c$
(Fig. \ref{fig3dst2} b), and strongly asymmetric at criticality (Fig. \ref{fig3dstc} b).

\section{ Finite-size scaling in the critical region  }

\label{fss}

For Anderson transitions, finite-size scaling involves
the multifractal spectrum at criticality
but a single correlation length exponent $\nu$ (see the reviews \cite{Jan,Huck}).
In this section, we thus try
the following  finite-size scaling form in the
critical region 
\begin{equation}
\overline{ Y_q(L,T) } = L^{- \tilde \tau (q) } \  \Phi \left( (T-T_c)
L^{1/\nu} \right)
\label{fssyq}
\end{equation}

\begin{figure}[htbp]
\includegraphics[height=6cm]{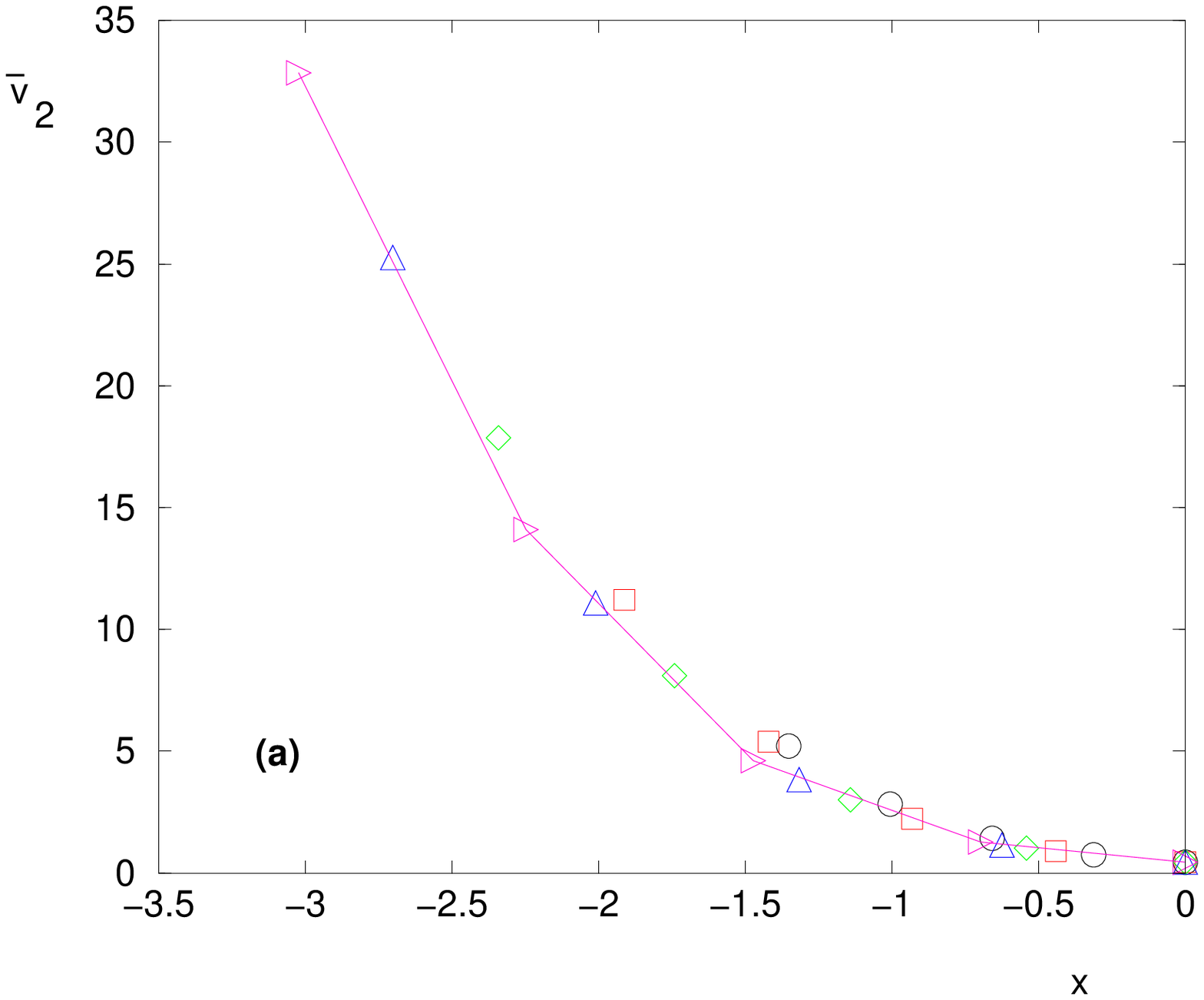}
\hspace{1cm}
\includegraphics[height=6cm]{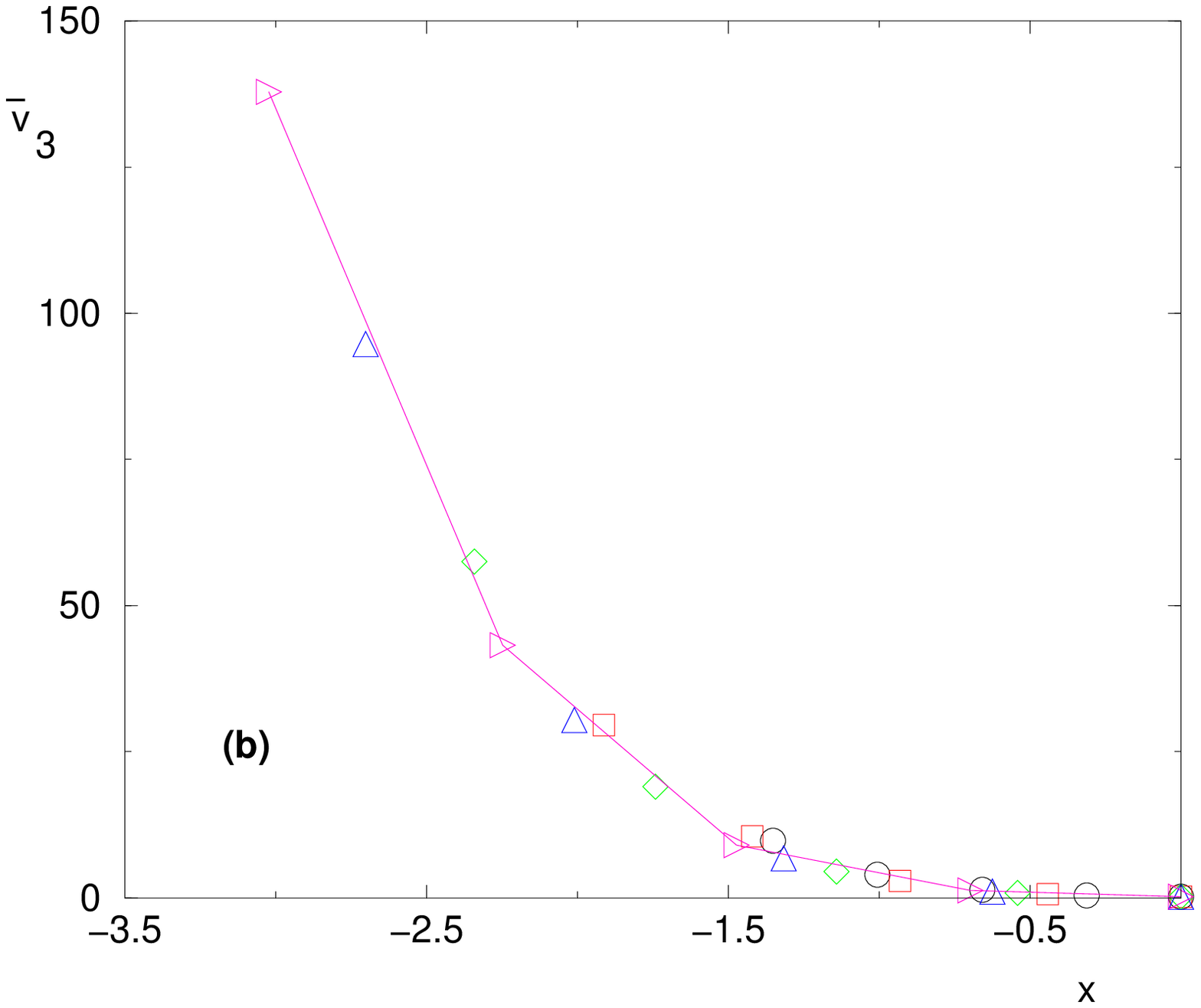}
\caption{(Color online) Finite-size scaling ($T<T_c$) of
$\overline{v_q}=L^{\tilde \tau(q)} \overline{Y_q}$ as a function of
$x=(T-T_c)L^{1/\nu}$, see Eq. \ref{fssyq}, with the value $\nu=2$ and
for the sizes $L=12 (\bigcirc),24 (\square),36 (\Diamond),48
(\triangle),60 (\rhd)$
(a)  $q=2$ (b) $q=3$
 }
\label{figfss}
\end{figure}

For $T<T_c$, the convergence to finite-values $\overline{
Y_q(L=\infty,T)}$  in the $L \to \infty$ limit yields 
\begin{equation}
\overline{ Y_q(L=\infty,T) } = (T_c-T)^{ \tilde \beta(q) } \ \ \ {\rm with } \
\ \tilde \beta(q)=\nu \tilde \tau (q)
\end{equation}
This relation between the multifractal exponents $\tilde \tau (q)$ and
the critical exponents $\tilde \beta(q)$ and $\nu$ is well known for the Anderson
transitions (see the reviews \cite{Jan,Huck}).
Our results for $T<T_c$ are shown on Fig. \ref{figfss} for $q=2$ (a)
and $q=3$ (b) with the value $\nu =2$.
This value is one of the two values $\nu \sim 2$ and $\nu \sim 4$
found previously in the literature for other observables \cite{Der_Gol,Ki_Br_Mo,DPtransi3d}.

\begin{figure}[htbp]
\includegraphics[height=6cm]{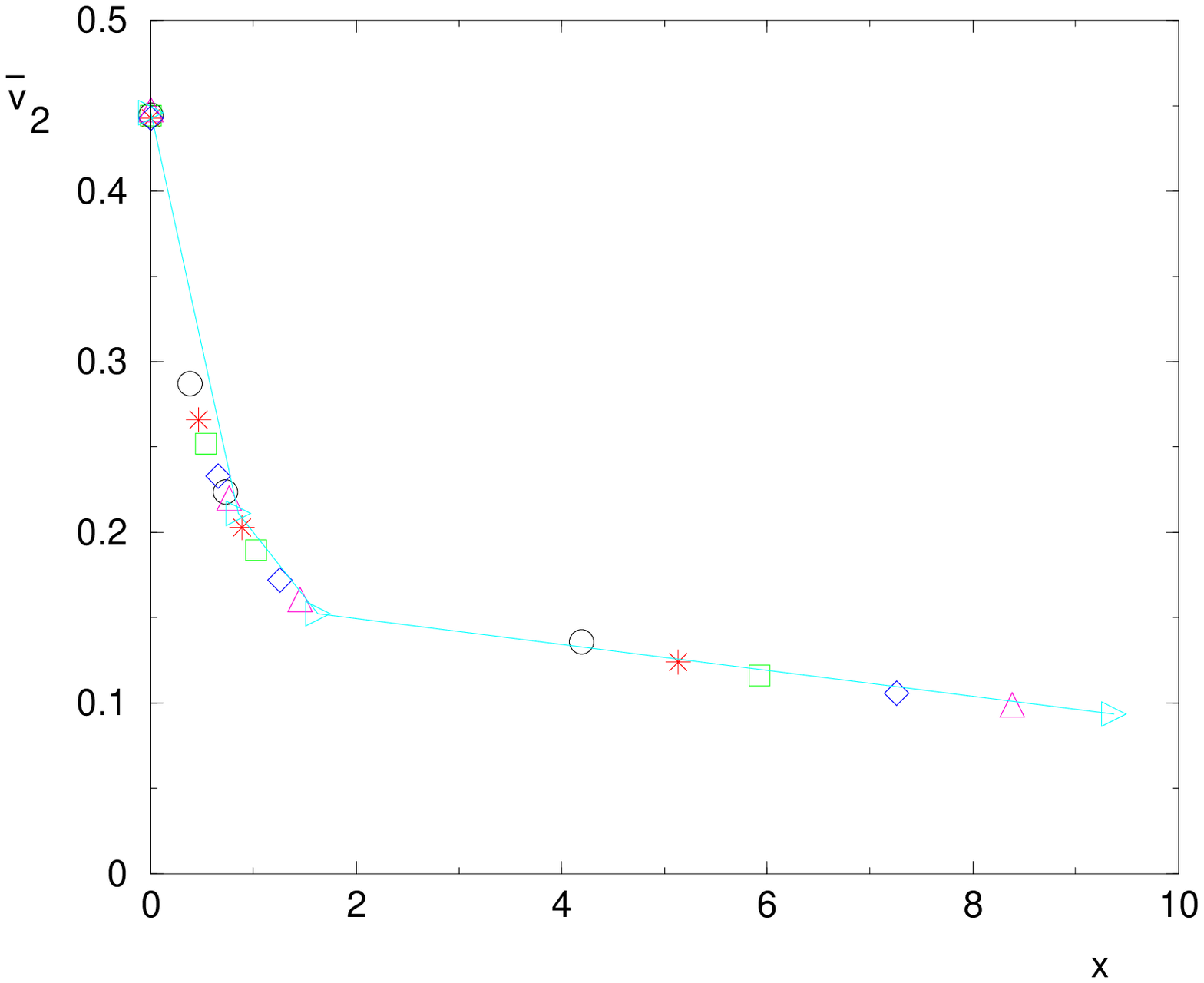}
\hspace{1cm}
\caption{(Color online)
  Finite-size scaling ($T>T_c$) of $\overline{v_2}=L^{\tilde
\tau(2)} \overline{Y_2}$ as a function of 
$x=(T-T_c)L^{1/\nu}$ with the value $\nu=2$ and for the sizes $L=12
(\bigcirc),18 (\ast),24 (\square),36 (\Diamond),48 
(\triangle),60 (\rhd)$, see Eq. \ref{fssyq}.}
\label{figfssabove}
\end{figure}

For $T>T_c$, the results of the finite-size scaling form of
Eq. \ref{fssyq} are shown on Fig. \ref{figfssabove} for $q=2$ with the
value $\nu=2$.
Note that the matching between the finite-size scaling form of Eq. \ref{fssyq}
and the asymptotic behavior in the delocalized phase
(Eq. \ref{deloc}) yields a diverging amplitude $A(T)$ in Eq. \ref{deloc} for $q>1$
 \begin{equation}
 \overline{ Y_q(L,T > T_c) } \sim  \frac{ A(T)} 
 { L^{ \frac{3}{2} (q-1)} } \ \ {\rm with} \ \ A(T) \sim 
  (T-T_c)^{- \nu (q-1) ( \frac{3}{2}- \tilde D(q)) }
 \label{fssyqaboveasymp}
 \end{equation}

Our data thus points towards a correlation length exponent $\nu \sim
2$ above and below $T_c$, i.e. towards a value very close to the
general lower
bound $\nu \geq 2/d=2$ of disordered systems \cite{chayes}.

\section{Conclusion} 

\label{conclusion} 

In this paper, we have found that the directed polymer in a random medium of dimension $1+3$
exhibits multifractal properties at the critical localization/delocalization transition.
We have numerically studied the statistics of the $Y_q(L)$
(see Eqs.\ref{defw} and \ref{ykdef}),
which are the dynamical analogs
of the Inverse Participation Ratios of Anderson localization quantum
models \cite{Jan,Huck,Mirlin_Evers}. Our results are
very close to the Evers-Mirlin scenario \cite{Mirlin_Evers,Mirlin2002}
for the Anderson transitions case. In particular, we have found that
the generalized dimensions $D(q)$ and $\tilde D(q)$ for typical and disorder averaged
values coincide for $q <q_c \sim 2$ but differ for $q>q_c$,
and that the probability distributions of $y=Y_q(L)/Y^{typ}_q(L) $
over the samples becomes scale invariant with a power-law tail $1/y^{1+x_q}$.
We have also measured the corresponding typical singularity spectrum $f(\alpha)$, which starts at
the value $\alpha_{min} =D(+\infty) \sim 0.77$, and ends at
$\alpha_{max}=+\infty$. Off-critical results lead, through a finite size
scaling analysis, to a value $\nu \sim 2$ for the correlation length 
exponent on both sides of the transition.

Finally, our numerical results, in particular the scale
invariant shape of the histogram of $\ln Y_2(L)$ shown on
Fig. \ref{fighistolny2} (a), strongly support the equality
$T_c=T_2(d=3)$ (see Eq. \ref{tc3d} and the corresponding discussion).

Since the directed polymer in a
random medium can be mapped onto a growth model in the
Kardar-Parisi-Zhang universality class\cite{Hal_Zha},
multifractality is also expected to show up at the critical point
of these growth models.
More generally, the present study confirms that it may be interesting
to characterize the critical points of quenched disordered models
by their multifractal properties.

\section*{Acknowledgements}

We thank F. Igl\'oi for drawing our attention to many relevant
references.


\begin{thebibliography}{99}

\bibitem{halsey}
T.C. Halsey, M.H. Jensen, L.P. Kadanoff, I. Procaria and B. Shraiman,
Phys. Rev. A 33, 1141 (1986).

\bibitem{Pal_Vul}
G. Paladin and A. Vulpiani, Phys. Rep. 156, 147 (1987).

\bibitem{Stan_Mea}
H.E. Stanley and P. Meakin, Nature 335, 405 (1988).

\bibitem{Aha}
A. Aharony and J. Feder Eds, {\it Fractals in Physics}, Essays in
honour of B.B. Mandelbrot, North Holland (1990).

\bibitem{Meakin}
P. Meakin, {\it Fractals, scaling and growth far from equilibrium},
Cambridge (1998).

\bibitem{harte}
D. Harte, '' Multifractals, Theory and Applications'', Chapman and Hall (2001).


\bibitem{duplantier_houches}
 B. Duplantier, ''Conformal Random Geometry'',
 Les Houches, Session LXXXIII, 2005, Mathematical Statistical Physics,
Eds A. Bovier et al., 101, Elsevier  (2006).





\bibitem{Ludwig}
A.W.W. Ludwig, Nucl. Phys. B 330, 639 (1990).




\bibitem{Jac_Car}
J.L. Jacobsen and J.L. Cardy, Nucl. Phys., B515, 701 (1998).

\bibitem{Ols_You}
T. Olsson and A.P. Young, Phys. Rev., B60, 3428 (1999).

\bibitem{PCBI}
G. Pal\'agyi, C. Chatelain, B, Berche and F. Igl\'oi, Eur. Phys. J,
B13, 357 (2000).

\bibitem{Cha_Ber}
C. Chatelain and B. Berche, Nucl. Phys., B572, 626 (2000).


\bibitem{Sourlas}
N. Sourlas, Europhys. Lett. 3, 1007 (1987).

\bibitem{Thi_Hil}
M.J. Thill and H.J. Hilhorst, J. Phys. I 6, 67 (1996).

\bibitem{Par_Sou}
G. Parisi and N. Sourlas, Phys. Rev. Lett. 89, 257204 (2002).

\bibitem{multiscaling}
J. Kisker and A. P. Young
Phys. Rev. B 58, 14397-14400 (1998) ;
F. Igloi, R. Juhasz, and H. Rieger
Phys. Rev. B 61 11552 (2000).

\bibitem{Weg}
F. Wegner, Z. Phys. B 36, 209 (1980).

\bibitem{Cas_Pel}
C. Castellani and L. Peliti, J. Phys. A 19, L429 (1986)

\bibitem{Schreiber}
M. Schreiber and H. Grussbach, Phys. Rev. Lett. 67, 607 (1991);
H. Grussbach and M. Schreiber, Phys. Rev. B 51, 663 (1995).


\bibitem{Terao}
T. Terao, Phys. Rev. B 56, 975 (1997).

\bibitem{Mirlin2002} A. Mildenberger, F. Evers, and A. D. Mirlin
Phys. Rev. B 66, 033109 (2002) 
 
\bibitem{Mirlin2006}
A. D. Mirlin, Y. V. Fyodorov, A. Mildenberger, and F. Evers
Phys. Rev. Lett. 97, 046803 (2006) 


\bibitem{Jan}
M. Janssen, Int. J. Mod. Phys. 8, 943 (1994);
M. Janssen, Phys. Rep. 295, 1 (1998).

\bibitem{Huck}
B. Huckestein, Rev. Mod. Phys. 67, 357 (1995).


\bibitem{Cha_etcie}
 C. Chamon, C. Mudry, and X.-G. Wen
Phys. Rev. Lett. 77, 4194-4197 (1996) ;
 H. E. Castillo, C. Chamon, E. Fradkin, P. M. Goldbart, C. Mudry,
 Phys. Rev. B56, 10668 (1997).

\bibitem{Mirlin_Evers}
F. Evers and A.D. Mirlin, Phys. Rev. Lett. 84 , 3690 (2000) ;
 A.D. Mirlin and F. Evers, Phys. Rev. B 62, 7920 (2000).
 

\bibitem{Pook}
W. Pook and M. Janssen, Z. Phys. B 82, 295 (1991)


\bibitem{timeexponents}
B. Huckestein and L. Sweitzer, Phys. Rev. Lett. 72, 713 (1994);
B. Huckestein and R. Klesse, Phys. Rev. B 55, R7303 (1997);
T. Brandes, B. Huckestein and L. Schweitzer,
Ann. Physk 5, 633 (1996).


\bibitem{averaged-typical}
F. Evers, A. Mildenberger and A.D. Mirlin, Phys. Rev. B 64, 241003
(2001).

\bibitem{Hal_Zha}
T. Halpin-Healy and Y.-C. Zhang, Phys. Repts., {\bf 254}, 215 (1995).


\bibitem{Der_Spo}
B. Derrida and H. Spohn, J. Stat. Phys., {\bf 51}, 817 (1988).


\bibitem{Mez}
M. M\'ezard, J. Phys. (France), {\bf 51}, 1831 (1990).

\bibitem{Car_Hu}
P. Carmona  and Y. Hu, Prob. Th. and Rel. Fields 124, 431 (2002);
P. Carmona  and Y. Hu, math.PR/0601670. 


\bibitem{Com}
F. Comets, T. Shiga and N. Yoshida, Bernoulli  9, no. 4, 705 (2003).

\bibitem{Var}
V. Vargas, math.PR/0603233.

\bibitem{Imb_Spe}
J. Z. Imbrie and T. Spencer, J. Stat. Phys. {\bf 52}, 609 (1988).


\bibitem{Coo_Der}
J. Cook and B. Derrida, J. Stat. Phys. {\bf 57}, 89 (1989).

\bibitem{Ki_Br_Mo}
J.M. Kim, A.J. Bray and M.A. Moore, Phys. Rev. {\bf A44}, R4782 (1991).




\bibitem{Der_Gol}
B. Derrida and O. Golinelli, Phys. Rev. {\bf A41}, 4160 (1990).

\bibitem{Der_Eva}
M. R. Evans and B. Derrida, J. Stat. Phys. 69 , 427 (1992).



\bibitem{DPcritidroplet}
C. Monthus and T. Garel,
Phys. Rev. E 74, 011101 (2006)


\bibitem{DPtransi3d}
C. Monthus and T. Garel, Eur. Phys. J. B 53, 39-45 (2006)




\bibitem{mandelbrot}
B. Mandelbrot, Physica A 163, 306 (1990)
B. Mandelbrot, J. Stat. Phys. 110, 739 (2003)

\bibitem{Chh_neg}
A.B. Chhabra and K.R. Sreenivasan, Phys. Rev. A 43, 1114 (1991).


\bibitem{Jen_neg}
M.H. Jensen, G. Paladin and A. Vulpiani, Phys. Rev. E 50 , 4352 (1994).

\bibitem{has_dup}
T.C. Halsey, K. Honda and B. Duplantier, J. Stat. Phys. 85, 681 (1996);
T.C. Halsey, B. Duplantier and K. Honda, Phys. Rev. Lett. 78, 1719 (1997). 




\bibitem{Chh}
A. Chhabra and R.V. Jensen, Phys. Rev. Lett. 62, 1327 (1989).

\bibitem{chayes}
 J. T. Chayes, L. Chayes, D.S. Fisher and T. Spencer
 Phys. Rev. Lett. {\bf 57}, 2999 (1986).






\end{thebibliography}
\end{document}